\documentclass[aps,prl,twocolumn,showpacs,superscriptaddress,floatfix]{revtex4}

\usepackage{graphicx}
\usepackage{dcolumn}
\usepackage{amsmath}
\usepackage{amsfonts}
\usepackage{amssymb}
\usepackage{bm}
\usepackage{color}
\usepackage{ulem}
\usepackage[colorlinks=true,linkcolor=blue,citecolor=blue,pdfauthor=Gross]{hyperref}

\begin{document}

\preprint{M.~Mariantoni {\it et al.}, version: 15 March 2010}

\title{Planck Spectroscopy and the Quantum Noise of Microwave Beam Splitters}

\author{M. Mariantoni}
\thanks{present address: Department of Physics, University of
California, Santa Barbara, California 93106, USA}
 \email{matmar@physics.ucsb.edu}
 \affiliation{Walther-Mei{\ss}ner-Institut, Bayerische Akademie der
              Wissenschaften, D-85748 Garching, Germany}

 \affiliation{Physik-Department, Technische Universit\"{a}t
              M\"{u}nchen, D-85748 Garching, Germany}

\author{E.~P.~Menzel}
 \affiliation{Walther-Mei{\ss}ner-Institut, Bayerische Akademie der
              Wissenschaften, D-85748 Garching, Germany}

\author{F.~Deppe}
 \affiliation{Walther-Mei{\ss}ner-Institut, Bayerische Akademie der
              Wissenschaften, D-85748 Garching, Germany}

 \affiliation{Physik-Department, Technische Universit\"{a}t
              M\"{u}nchen, D-85748 Garching, Germany}

\author{M.~\'{A}.~Araque~Caballero}
 \affiliation{Walther-Mei{\ss}ner-Institut, Bayerische Akademie der
              Wissenschaften, D-85748 Garching, Germany}

 \affiliation{Physik-Department, Technische Universit\"{a}t
              M\"{u}nchen, D-85748 Garching, Germany}

\author{A.~Baust}
 \affiliation{Walther-Mei{\ss}ner-Institut, Bayerische Akademie der
              Wissenschaften, D-85748 Garching, Germany}

 \affiliation{Physik-Department, Technische Universit\"{a}t
              M\"{u}nchen, D-85748 Garching, Germany}

 \affiliation{Physik-Department, Technische Universit\"{a}t
              M\"{u}nchen, D-85748 Garching, Germany}

\author{T.~Niemzcyk}
 \affiliation{Walther-Mei{\ss}ner-Institut, Bayerische Akademie der
              Wissenschaften, D-85748 Garching, Germany}

 \affiliation{Physik-Department, Technische Universit\"{a}t
              M\"{u}nchen, D-85748 Garching, Germany}

\author{E.~Hoffmann}
 \affiliation{Walther-Mei{\ss}ner-Institut, Bayerische Akademie der
              Wissenschaften, D-85748 Garching, Germany}

 \affiliation{Physik-Department, Technische Universit\"{a}t
              M\"{u}nchen, D-85748 Garching, Germany}

\author{E.~Solano}
 \affiliation{Departamento de Qu\'{i}mica F\'{i}sica, Universidad del Pa\'{i}s
              Vasco–-Euskal Herriko Unibertsitatea, 48080 Bilbao, Spain}

 \affiliation{IKERBASQUE, Basque Foundation for Science, Alameda Urquijo 36, 48011 Bilbao, Spain}

\author{A.~Marx}
 \affiliation{Walther-Mei{\ss}ner-Institut, Bayerische Akademie der
              Wissenschaften, D-85748 Garching, Germany}

\author{R.~Gross}
 \email{Rudolf.Gross@wmi.badw.de}
 \affiliation{Walther-Mei{\ss}ner-Institut, Bayerische Akademie der
              Wissenschaften, D-85748 Garching, Germany}

 \affiliation{Physik-Department, Technische Universit\"{a}t
              M\"{u}nchen, D-85748 Garching, Germany}

\begin{abstract}
We use a correlation function analysis of the field quadratures to characterize
both the black body radiation emitted by a $50\,\Omega$ load resistor and the
quantum properties of two types of beam splitters in the microwave regime. 
To this end, we first study vacuum fluctuations as a function of frequency 
in a Planck spectroscopy experiment and then measure the covariance
matrix of weak thermal states. Our results provide direct experimental evidence that vacuum fluctuations represent the fundamental minimum quantum noise added by a beam
splitter to any given input signal.
\end{abstract}

\date{\today}


\maketitle

At optical frequencies, single-photon detectors~\cite{mandel:1995:a} and beam splitters 
are key ingredients for the successful development of atomic physics and quantum 
optics. These devices are crucial for the implementation of quantum homodyne 
tomography~\cite{leonhardt:1997:a}, quantum information processing and 
communication~\cite{bouwmeester:2000:a}, as well as all-optical quantum 
computing~\cite{kok07}. The recent advent of circuit quantum electrodynamics 
(QED)~\cite{wallraff:2004:a,schoelkopf:2008:a,blais:2004:a,astafiev:2007:a,sillanpaa:2007:a,majer:2007:a,houck:2007:a,deppe:2008:a,mariantoni:2008:a,niemczyk:2009:a,bozyigit:2010:a}
has paved the way for the generation of single photons in the microwave (mw)
regime~\cite{houck:2007:a,bozyigit:2010:a}. Despite the rapid advances in
this prospering field, the availability of suitable mw 
photodetectors~\cite{helmer:2009:a,romero:2009:a} and well-characterized mw
beam splitters is still at an early stage. However, we have recently shown that the use 
of low-noise cryogenic high electron mobility transistor (HEMT) amplifiers represents a 
versatile and powerful approach for the analysis of both classical and non-classical mw
signals on a single photon level. Although the phase-insensitive HEMT amplifiers 
obscure the signal by adding random noise of typically 10--20
photons at 5\,GHz, they do not perturb the correlations of signals opportunely
split into two parts and then processed by two parallel amplification and
detection chains. In such a setup, a correlation analysis allows for full state 
tomography of propagating quantum mw signals and the detector noise, 
simultaneously~\cite{menzel:2010:a}. We note that HEMT amplifiers represent available
``off-the-shelf'' technology and offer flat gain over a broad frequency range
of several GHz. Here, we present results of two experiments demonstrating the 
successful application of our setup to the characterization of weak thermal states. In 
a first experiment denoted as Planck spectroscopy we analyze the mw black body
radiation emitted by a matched $50\,\Omega$ load resistor as a function of
temperature in the frequency regime $4.7\le \omega/2\pi
\le 7.1$\,GHz. Besides confirming that the mean thermal photon number follows
Bose-Einstein statistics~\cite{gabelli:2004:a,movshovich:1990:a}, our data
directly show that the quantum crossover temperature $T_{\rm cr}$ shifts with
frequency as $T_{\rm cr} = \hbar\omega/2k_B$, providing an indirect measure of
mw vacuum fluctuations with high fidelity. In a second experiment, we use weak
thermal states for a detailed experimental characterization of mw beam
splitters at the quantum level. This task, which has not been accomplished
previously, is particularly important because mw beam splitters are key
elements in a variety of quantum-optical experiments such as Mach-Zehnder and
Hanbury Brown and Twiss interferometry~\cite{mandel:1995:a,gabelli:2004:a}.

\begin{figure}[tb]
\centering{
 \includegraphics[clip=,width=0.9\columnwidth]{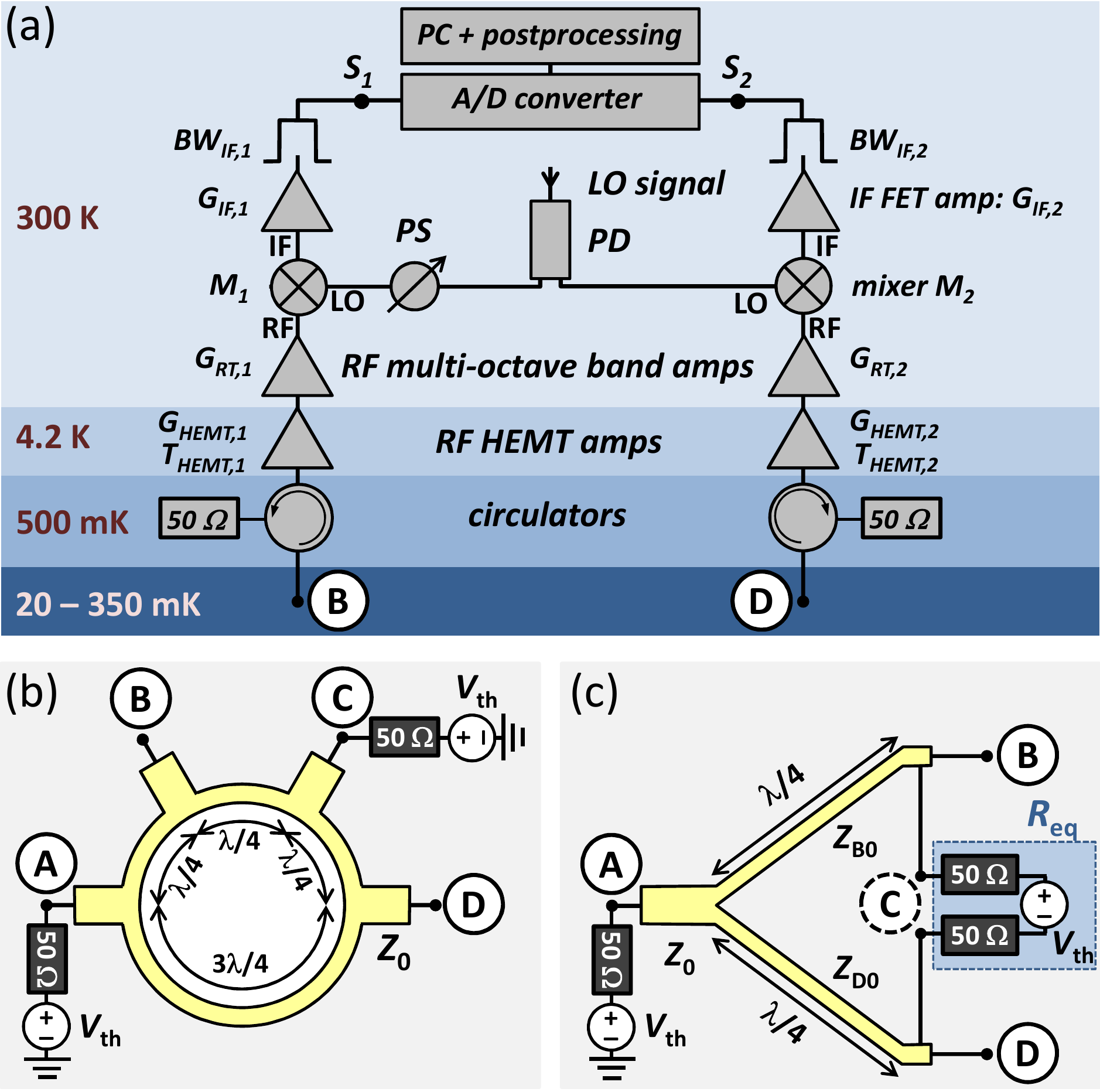}}
\caption{
Schematics of the experimental setup (cf.~main text). (a) Amplification and
detection chains. (b) $180^{\circ}$ hybrid ring (HR). (c) Wilkinson power
divider (WPD). For the WPD, port C represents a hidden internal port and $R_{\rm
eq}$ an internal distributed resistor, which can be modeled as two
equivalent matched $50\,\Omega$ loads adding correlated thermal noise via the
hidden port C only.}
 \label{fig1:setup:matteo:mariantoni}
\end{figure}

The experimental setup is shown in Fig.~\ref{fig1:setup:matteo:mariantoni}.
As an ideal black-body source emitting thermal mw states~\cite{beenacker:2001:a}, we 
use matched $50\,\Omega$ loads whose temperature $T$ can be varied between 20 and 
350\,mK and measured with a RuO thermometer. The temperature is
measured by a RuO thermometer. The associated quantum voltage can be expressed
as $\hat{V}_{\rm th} = V_0 (\hat{p}^{\dag} + \hat{p})$, where $V_0^2 =
4\textrm{BW} R_0 \hbar\omega/2$, $R_0=50\,\Omega$, $\hat{p}^{\dag}$ and
$\hat{p}$ are bosonic creation and annihilation operators, and BW is the
bandwidth. The thermal mw signal is fed to the input ports of a 3\,dB mw beam
splitter. We perform experiments on two different beam splitter realizations: a
four-port $180^{\circ}$ hybrid ring (HR)
[cf.~Fig.~\ref{fig1:setup:matteo:mariantoni}(b)] and a three-port Wilkinson
power divider (WPD) [cf.~Fig.~\ref{fig1:setup:matteo:mariantoni}(c)]. For the
WPD, an internal distributed resistor $R_{\rm eq}$ shunting the output ports B
and D provides isolation between those ports and impedance matching for port A.
In addition, an external $50\,\Omega$ load is attached to input port A. For the
HR, matched $50\,\Omega$ loads are attached to both input ports A and C. The
input/output relations of the HR are $\hat{V}_{\rm B}= -\imath \, (\hat{V}_{\rm
A} + \hat{V}_{\rm C}) / \sqrt{2}$ and $\hat{V}_{\rm D} = -\imath \,
(-\hat{V}_{\rm A} + \hat{V}_{\rm C}) / \sqrt{2}$~\cite{collin:2000:a}. Remark
that the WPD, although appearing to be a three-port device, has to be treated
quantum mechanically as having an additional ``hidden'' internal fourth port C,
see Fig.~\ref{fig1:setup:matteo:mariantoni}(c), ensuring energy conservation
and commutation relations. In this case the input/output relations are
$\hat{V}_{\rm B}= -\imath \, (\hat{V}_{\rm A} - \hat{V}_{\rm C})/\sqrt{2}$ and
$\hat{V}_{\rm D}= -\imath \, (\hat{V}_{\rm A} + \hat{V}_{\rm C})/\sqrt{2}$.
Regarding thermal noise, the internal resistor $R_{\rm eq}$ can be modeled as
two equivalent matched $50\,\Omega$ loads adding correlated thermal noise via
the hidden port C only. If the input signal at port A is large, this additional
noise can be neglected and the WPD can be formally treated as a three-port
device.

As shown in Fig.~\ref{fig1:setup:matteo:mariantoni}(a), the output ports B and
D of the beam splitter are connected to two symmetric amplification and
detection chains, each comprising the following elements: (i) A cryogenic
circulator at $500$\,mK with 21\,dB isolation which prevents amplifier noise from 
leaking back to the mw source under study. Furthermore,
together with the isolation of two outputs of the beam splitters they avoid
spurious correlations of the noise originating from the two amplification
chains. (ii) A low-noise HEMT amplifier thermally anchored at $4.2$\,K with
power gain $G_{\rm HEMT} \simeq 24$\,dB and noise temperature $T_{\rm
HEMT}\simeq 6 \pm 1$\,K. The noise added by the linear HEMT amplifiers  is the
dominating amplifier noise in each detection channel and is expressed by
$\hat{\chi} = V_0 (\hat{\xi}^{\dag} + \hat{\xi})$, where $\hat{\xi}^{\dag}$ and
$\hat{\xi}$ are bosonic creation and annihilation
operators~\cite{caves:1982:a}. (iii) A room-temperature amplifier and (iv) a
mixer (M) and local oscillator (LO) to down convert the mw signal to the
intermediate frequency (IF). The phase $\varphi_{\rm LO}$ of one of the two LO
signals obtained by means of a power divider (PD) can be varied by a phase
shifter (PS). (v) An IF amplifier and (vi) a 7\,kHz to 26\,MHz band pass filter. 
Using a double sideband receiver the total bandwidth BW in our experiment is twice 
the bandwidth of the bandpass filter. The down converted voltage signal
components $s_1$ and $s_2$ in the two chains are
finally synchronously digitized by an acquisition board with $4\times
10^8$\,samples/s and 12\,bit resolution.

\begin{figure*}[tb]
\centering{
 \includegraphics[clip=,width=0.9\textwidth]{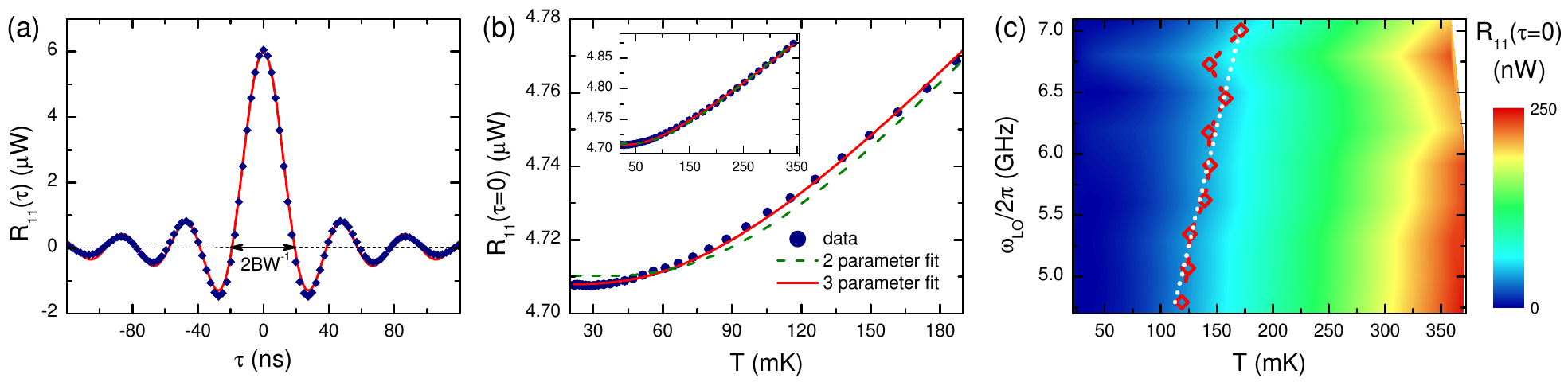}}
\caption{Planck spectroscopy of thermal mw states at  $\omega/2\pi = 5.3$\,GHz
using a WPD. (a) Auto-correlation function $R_{11} (\tau)$. The symbols
represent the data, the line a fitted curve. (b) Temperature dependence of the
variance $R_{11} (\tau{=}0)$ (Planck function). The symbols represent the data,
the dotted and full lines are obtained by two and three parameter fits,
respectively. The inset shows the data over a wider $T$ range. (c) Planck
spectroscopy. Contour plot of $R_{11}(\tau=0)$ versus $T$ for different $\omega
/2\pi$. The data has been corrected for the frequency dependent amplifier gain,
the offset due to vacuum and amplifier noise has been subtracted. The symbols and 
the dashed line mark the experimentally measured and theoretically expected 
quantum crossover temperatures, respectively. }
 \label{fig2:planck:matteo:mariantoni}
\end{figure*}

According to Nyquist~\cite{nyquist:1928:a}, the black body radiation emitted by
a resistor with resistance $R_0$ within the bandwidth BW around the frequency
$\omega=\omega_\text{LO}$ is given by $\langle V_{\rm th}^2 \rangle /R_0 = 
4\mathrm{BW} \langle n_{\rm th} \rangle \hbar\omega$. Here, $\langle n_{\rm th} 
\rangle$ is the
average thermal photon population. It has been shown that for conductors with a
large number of electronic modes the statistics of the emitted photons is given
by a Bose-Einstein distribution~\cite{beenacker:2001:a}. In this case the well
known result $\langle V_{\rm th}^2\rangle /R_0 = 4\mathrm{BW}
\frac{\hbar\omega}{2} \coth (\hbar\omega / 2k_{\rm B}T)$ is obtained. The power
emitted into a perfectly matched circuit with characteristic impedance $Z_0$ is
reduced by a factor of $1/4$ due to voltage division. Together with the beam
splitter input/output relations of the HR, the signal components $s_1$ and
$s_2$ are given by $s_1 = \sqrt{G_1} (-\imath \alpha V^A_{\rm th} -\imath \beta
V^C_{\rm th} + \chi_1)$ and $s_2 = \sqrt{G_2} (\imath \alpha V^A_{\rm th} -
\imath \beta V^C_{\rm th} + \chi_2)$. Here, $G_1 \simeq G_2$ is the total power
gain of the amplification chains, $\alpha = \beta = 1/2\sqrt{2}$, and $\chi_1$
and $\chi_2$ are the independent noise contributions of the amplifiers.
Equivalently, for the WPD we obtain $s_1 = \sqrt{G_1} (-\imath\alpha V^A_{\rm
th} +\imath \beta V^C_{\rm th} + \chi_1)$ and $s_2 = \sqrt{G_2} (-\imath\alpha
V^A_{\rm th} -\imath \beta V^C_{\rm th} + \chi_2)$. We note that $V^A_{\rm
th}$, $V^C_{\rm th}$ and $\chi_{1,2}$ are classical realizations of the
operators given above. By recording a large number of $1\mu$s-long time traces 
($\sim 10^6$), the auto- and cross-correlation functions $R_{ii}
(\tau ) = \langle s^{\ast}_i ( t + \tau ) \, s_i (t) \rangle /Z_0 =
\sigma^2_{ii} \, {\rm sinc} ({\rm BW}\,\tau) /Z_0$ and $R_{ij} (\tau) = \langle
s^{\ast}_i (t +\tau) \, s_j (t) \rangle /Z_0 = \sigma^2_{ij} \, {\rm sinc}
({\rm BW}\, \tau) \, \cos\varphi_{\rm LO} /Z_0$, respectively, can be
calculated ($i,j=1,2$). Since $\langle s_i (t) \rangle = 0$ for thermal states,
the auto- and cross-correlation functions are equal to the auto-variance
$C_{ii}(\tau) = R_{ii}(\tau) - \langle s_i \rangle^2$ and cross-variance
$C_{ij}(\tau) = R_{ij}(\tau) - \langle s_i \rangle\langle s_j \rangle$,
respectively. Here, $\tau$ is the time shift between two traces being
correlated, $\sigma^2_{ii}$ and $\sigma^2_{ij}$ are the variance and covariance
of the voltage signals $s_1$ and $s_2$.

\begin{figure*}[tb]
\centering{
 \includegraphics[clip=,width=0.9\textwidth]{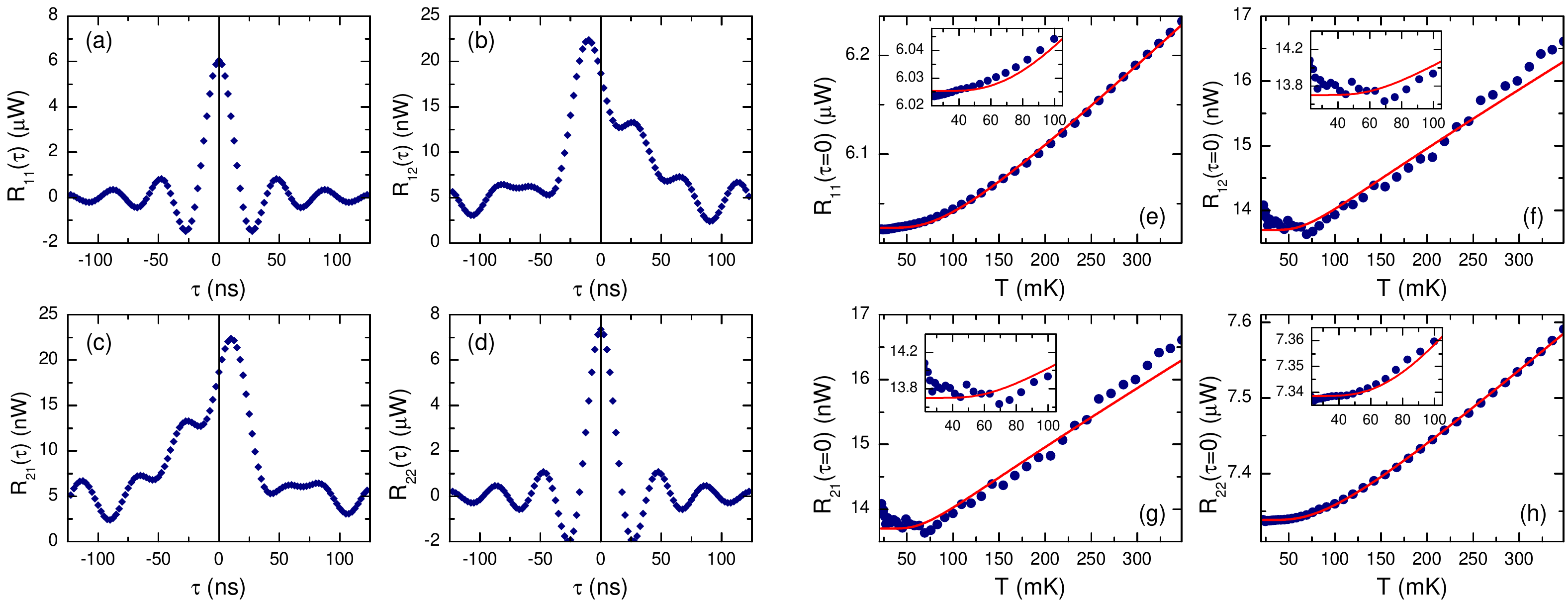}}
\caption{Full correlation function and covariance matrices measured at $\omega
/2\pi = 5.0$\,GHz using a WPD ($G_1\simeq 90.5$\,dB, $G_2\simeq 91.3$\,dB, and
$G_{12}\simeq 90.9$\,dB. (a)--(d) $R_{ii}(\tau)$ and $R_{ij}(\tau)$ measured at
$T=30$\,mK. (e)--(h) Temperature dependence of $R_{ii}(0)$ and $R_{ij}(0)$. The
symbols represent experimental data, the lines fits to the data.}
 \label{fig3:correlation:matteo:mariantoni}
\end{figure*}

We first discuss the Planck spectroscopy experiment. Here, we use only a single
amplification chain and determine the auto-correlation function $R_{11}(\tau)$
or $R_{22}(\tau)$. Exemplary, Fig.~\ref{fig2:planck:matteo:mariantoni}(a) shows
the measured $R_{11}(\tau)$ curve obtained for $T=30$\,mK using a WPD. A very
similar result is obtained for $R_{22}(\tau)$ [see
Fig.~\ref{fig3:correlation:matteo:mariantoni}(d)]. Fitting the data to $C_{11}
(\tau)$ allows us to extract the measurement bandwidth to ${\rm BW} \simeq
52$\,MHz. Assuming that the signal contributions $V^A_{\rm th}$ and $V^C_{\rm
th}$ due to the two load resistors and the noise $\chi_i$ of the HEMT amplifier
are independent, we simply can add up their variances and obtain $R_{ii}(0) =
C_{ii}(0) = \sigma^2_{ii} = \langle s_i^2\rangle /Z_0 = \frac{G_i}{Z_0} \left[
(\alpha^2+\beta^2) \langle V^2_{\rm th} \rangle + \langle \chi^2_i \rangle
\right]$. With $\hbar\omega/2k_{\rm B}T_{\rm HEMT} \ll 1$ we can introduce a
classical noise temperature $T_{\rm i,HEMT}$ for the amplifiers and obtain
\begin{eqnarray}
R_{ii}(0) & = &  \; G^\star_i \; {\rm BW} \;
\left[  \frac{\hbar\omega}{2} \coth \frac{\hbar \omega}{2
k_{\rm B} T} + k_{\rm B} T^\star_{\rm i,HEMT} \right]
 \, .
 \label{R:kk:0}
\end{eqnarray}
Here, $G^\star_i = \gamma G_i$ is the effective gain with $G_i$ the total gain
of the amplification chain and $T^\star_{\rm i,HEMT} = T_{\rm i,HEMT}/\gamma$
is the amplifier effective noise temperature, representing the amplifier noise
temperature relative to the input of the WPD. For both, the HR and WPD, $\gamma
= 4(\alpha^2 + \beta^2 ) =1$.

Figure~\ref{fig2:planck:matteo:mariantoni}(b) shows the measured variance
$R_{11}(0)$ as a function of $T$ for $\omega /2\pi = 5.3$\,GHz in the case of a
WPD. During the experiments, $T$ was varied between approximately $20$ and
$350$\,mK by means of a resistive heater and continuously monitored with a RuO
thermometer. As expected, the measured variance is close to a Planck function
and nicely reproduces the quantum crossover at $T_{\rm cr} = \hbar\omega/2k_B$.
A fit of Eq.~(\ref{R:kk:0}) to the data using $G^\star_1$ and $T^\star_{\rm
1,HEMT}$ as free parameters yields $G^\star_1 \simeq 90$\,dB and $T^\star_{\rm
1,HEMT} \simeq 6$\,K. Evidently, there are slight deviations between the data
and the two-parameter fit. This can be understood by taking into account that
the effective electronic temperature $T_{\rm eff}$ of the load resistors at
ports A and C may differ by a small amount $\delta T$. Using $\delta T$ as the
third fitting parameter the solid line in
Fig.~\ref{fig2:planck:matteo:mariantoni}(b) is obtained, demonstrating
excellent agreement with the experimental data. The $\delta T$ values obtained
by fitting the data are reasonably small and typically range between $1$ and
$10$\,mK. The large bandwidth of the HEMT amplifier allows us to perform
equivalent measurements at several frequencies between $4.7$ and $7.1$\,GHz.
The result of such Planck spectroscopy is shown in
Fig.~\ref{fig2:planck:matteo:mariantoni}(c). Clearly the quantum crossover
temperature $T_{\rm cr} $ shifts to higher values with increasing frequency.
Due to the finite uncertainty in $T_{\rm eff}$, we derive an effective
crossover temperature $T_{\rm cr} +\delta T_{\rm cr}$ [cf.~red crosses
connected by a red line in Fig.~\ref{fig2:planck:matteo:mariantoni}(c)] which
again slightly deviates from the theoretically expected value $T_{\rm cr} =
\hbar\omega/2k_{\rm B}$ [cf.~dashed white line in
Fig.~\ref{fig2:planck:matteo:mariantoni}(c)]. The magnitude $\delta T_{\rm cr}$
quantifies the measurement fidelity $\mathcal{F} \equiv 1 - | \delta T_{\rm cr}
| / T_{\rm cr}$ of our setup for vacuum fluctuations. Notably, for the entire
frequency range $\mathcal{F} \gtrsim 95$\,\%. In summary, our Planck
spectroscopy experiments not only provide clear evidence for the Bose-Einstein
statistics of photons emitted by a conductor in the few photon limit, but also
directly demonstrate the frequency dependence of the quantum crossover
temperature.

We next turn to the analysis of the mw beam splitters.
Figures~\ref{fig3:correlation:matteo:mariantoni}(a)-\ref{fig3:correlation:matteo:mariantoni}(d)
show the entire correlation matrix. The off-diagonal elements are
cross-correlation functions measured choosing $\varphi_{\rm LO}$ in order to
obtain a maximum positive result. This guarantees that the signals associated
with the two detection channels are skewed in phase and no unwanted
de-correlation is introduced during the experiments. Since the signal
contributions of the thermal noise sources and the amplifier noise are
independent, all cross-correlations vanish, e.g. $\langle\hat{\chi}_1
\hat{\chi}_2\rangle = \langle\hat{\chi}_1\rangle\langle \hat{\chi}_2\rangle =
0$. Then, for $\alpha^2 = \beta^2=1/8$ the covariance $R_{12}(0) = C_{12}(0) =
\sigma_{12}^2$ is obtained to
\begin{eqnarray}
R_{12}(0) & = & \frac{\hbar\omega}{4} G_{12} {\rm BW}
\left[ \coth \frac{\hbar\omega}{2k_{\rm B}T_{\rm A}}
     - \coth \frac{\hbar\omega}{2k_{\rm B}T_{\rm C}} \right]
 \label{R:12:0}
\end{eqnarray}
with the power cogain $G_{12}=\sqrt{G_1}\sqrt{G_2}$. We note that the
temperatures $T_{\rm A}$ and $T_{\rm C}$ of the load resistors at port A and C
are identical only in the ideal case, resulting in $R_{12}(0)=0$ for the HR and
WPD. However, in real experiments there may be small temperature differences
$\delta T = T_{\rm A} - T_{\rm C}$.
Figures~\ref{fig3:correlation:matteo:mariantoni}(e)-\ref{fig3:correlation:matteo:mariantoni}(h)
show the covariance matrix as a function of $T$ for $\omega /2\pi = 5.0$\,GHz
and using a WPD. The diagonal matrix elements $R_{11}(0)$ and $R_{22}(0)$
represent variance measurements and are analogous to the results of
Fig.~\ref{fig2:planck:matteo:mariantoni}(b). The off-diagonal elements,
instead, represent covariance measurements. It is evident that both the offset
signal at 20\,mK and the signal span between 20 and 350\,mK for the covariance
is reduced by about two orders of magnitude as compared to the variance. This
suggests that there is a cancellation of both the amplifier noise and the
signal when measuring the covariance. The former is due to the fact that the
amplifier noises are uncorrelated. The latter is expected from
Eq.~(\ref{R:12:0}). In order to prove this conjecture, we use
Eq.~(\ref{R:12:0}) to fit the experimental data. Doing so we use the cogain
which is determined by fitting the variance data. Furthermore, we set $T_{\rm
A}=T$, where $T$ is the temperature measured by the thermometer, and use
$\delta T$ as the free fitting parameter. In this way, we obtain the red curves
in Figs.~\ref{fig3:correlation:matteo:mariantoni}(f) and
\ref{fig3:correlation:matteo:mariantoni}(g), which are in excellent agreement
with the data. We obtain $\delta T$ values of a few mK. From the covariance
measurements we can get information on the characteristics of the WPD. Since
Eq.~(\ref{R:12:0}) explicitly assumes the existence of four ports, the perfect
fit of the experimental data provides clear evidence that the WPD effectively
behaves as a four-port device. In the quantum limit, this fourth port adds
vacuum noise to any given input signal.

\begin{figure}[b]
\centering{
 \includegraphics[clip=,width=0.9\columnwidth]{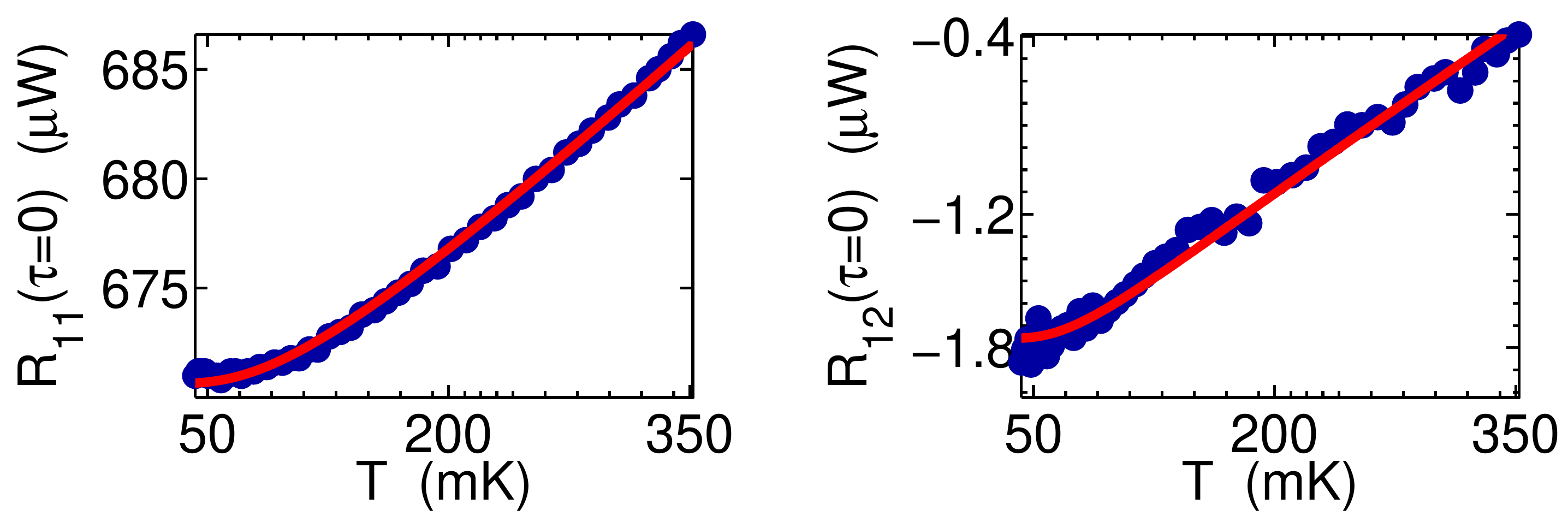}}
\caption{
Temperature dependence of the (a) variance $R_{ii}(0)$ and (b) covariance
$R_{ij}(0)$ measured at $\omega/2\pi =5.85$\,GHz using a HR. The different
power scale compared to Fig.~\ref{fig3:correlation:matteo:mariantoni}(e)-(h)
results from a different amplifier configuration.}
 \label{fig4:hr:matteo:mariantoni}
\end{figure}

In order to confirm our findings on the WPD, we have measured the $T$
dependence of the variance and covariance also for a HR
(cf.~Fig.~\ref{fig4:hr:matteo:mariantoni}), which is a straightforward
four-port beam splitter. The covariance data of
Fig.~\ref{fig4:hr:matteo:mariantoni}(b) are in excellent agreement with the
fitted curve obtained from Eq.~(\ref{R:12:0}). This clearly demonstrates that,
both the HR and WPD are characterized by the same fundamental
quantum-mechanical behavior.

In conclusion, we have applied a correlation function analysis of the field
quadratures to characterize black body radiation and the quantum properties of
mw beam splitters. Our Planck spectroscopy experiments show that the mean
thermal photon number emitted by a load resistor is following Bose-Einstein
statistics and that the quantum crossover temperature shifts with frequency as
$T_{\rm cr} = \hbar\omega/2k_B$, providing an indirect measure of mw vacuum
fluctuations with high fidelity. Moreover, we have shown that, at the quantum
level, vacuum fluctuations represent the minimum fundamental noise added by
both three- and four-port microwave beam splitters.

This work is supported by the German Research Foundation through SFB~631, the
German Excellence Initiative via NIM, NSERC, UPV-EHU GIU07/40, Ministerio de
Ciencia e Innovaci\'on FIS2009-12773-C02-01, EU EuroSQIP and SOLID projects.

\end{document}